\newcommand{\beq}{\begin{equation}}
\newcommand{\eeq}{\end{equation}}
\newcommand{\beqa}{\begin{eqnarray}}
\newcommand{\eeqa}{\end{eqnarray}}

\renewcommand{\lambda}{\ell}

\newcommand{\bS}{{\bf S}}
\newcommand{\bM}{{\bf M}}
\newcommand{\bk}{{\bf k}}
\newcommand{\br}{{\bf r}}
\documentstyle[aps,prl,twocolumn,epsf]{revtex}
 
\begin{document}
\def\dfrac#1#2{{\displaystyle{#1\over#2}}}

\twocolumn[\hsize\textwidth\columnwidth\hsize\csname @twocolumnfalse\endcsname

%\hfill{LA-UR-****}

\title{Spontaneous time reversal and parity breaking  in a
$d_{x^2-y^2}$-wave superconductor with magnetic impurities}

\author{
  A.V.~Balatsky  
}
\address{Theoretical Division, Los Alamos National Laboratory, Los Alamos, 
New Mexico 87545}
  
\date{\today}
\maketitle

\begin{abstract}
It is argued that the  $d_{x^2-y^2}$ superconductor
 exhibits the  transition into  $d_{x^2-y^2} + id_{xy}$ state
 in the presence of magnetic impurities in certain conditions. Linear coupling 
between impurity magnetization and the $d_{xy}$ component 
drives this transition. Violation of time-reversal symmetry 
and parity occurs spontaneously via the second order transition. 
In the ordered  phase both impurity magnetization and $d_{xy}$ 
component of the order parameter develop and are proportional to
 each other.  
 
\


\end{abstract}

\

\

]

%\newpage

It is well known that   magnetic impurities destroy the 
singlet superconducting 
state due to spin scattering which breaks pair singlets 
\cite{AG}. In the case 
of the gapless (with the nodes of the gap) d-wave
 superconductor both magnetic 
and nonmagnetic impurities produce a finite density 
of states at zero energy. 
These effects are a simple and direct consequence
 of the lifetime effects 
produced by impurities. These are well
 known ``incoherent'' effects of impurities 
in unconventional superconductors. After recent
experiments   by Movshovich et.al. 
\cite{Roman1}
 we are led to 
believe that  another phenomena is possible, namely
 the transition to  the 
second superconducting  phase as a result of condensate
 interactions with 
magnetic impurities. The time reversal  violating state
 is formed at low energy 
and the order parameter of the new phase is 
$d_{x^2-y^2} + id_{xy}$ ($d+id$). In this phase the impurity spins acquire 
nonzero spin density along z-axis, i.e. 
out of plane. 

The physical origin of the instability comes from the fact that the $d+id$ state
has an orbital moment which couples to the magnetic impurity spins. In the pure 
phase one can think of d-wave state as an equal admixture of the orbital moment 
$L_z = \pm 2$ pairs: 
\beqa
\Delta_0(\phi) = \Delta_0 \cos 2\phi ={\Delta_0\over{2}}(\exp(2i\phi) + 
\exp(-2i\phi))
\eeqa
Here $\phi$ is the 2D planar angle of the momentum on the Fermi surface, 
$\Delta_0$ is the magnitude of the $d_{x^2-y^2}$ component. We consider 2D 
$d_{x^2-y^2}$ superconductor, motivated by the layered   structure of the 
cuprates. In the presence of the (ferromagnetically) ordered impurity spins 
$S_z$ the coefficients of the $L_z = \pm 2$ components will shift {\em linearly} in $S_z$ with {\em opposite} signs:
\beqa
\Delta_0(\phi) \rightarrow {\Delta_0\over{2}}((1 + g S_z)\exp(2i\phi)\nonumber\\ 
+ (1 - g S_z)\exp(-2i\phi))
 = \Delta_0(\phi) + i \ S_z\Delta_1(\phi)
\eeqa
 where $\Delta_1(\phi) = g/2 \  \sin 2\phi$ -- is the $d_{xy}$ component and $g$ is the  coupling constant. The  relative phase $\pi/2$ of these two order 
parameters comes out naturally  because  $d+id$ state has a noncompensated 
orbital moment $L_z = +2$.  This  natural coupling of the orbital moment of the 
pairs to the spin  magnetic moment leads to  the instability of the  
$d_{x^2-y^2}$ state  towards time reversal violating $d+id$ state.

Recent experimental observation of the surface-induced time-reversal violating 
state in YBCO suggests that the secondary component of the order parameter 
(d+is) 
can be induced \cite{Green1}. Theoretical explanation, based on surface-induced 
 Andreev states has been suggested by Sauls and co-workers \cite{Sauls1}. The 
source of the secondary  component is the bending of the original  $d_{x^2-y^2}$ order parameter at the surface. 

In a different approach Laughlin \cite{Laugh1} argued that
 the $d_{x^2-y^2}$ state is unstable towards $d+id$ state 
in the bulk in the perpendicular magnetic field at low enough 
temperatures. The time reversal and parity are broken 
 by external field in this case. The linear coupling  of the
 secondary order parameter to the external field is central to
 his consideration and results in the first order phase transition
  into $d+id$ state. This transition was suggested to be responsible for the 
kink-like feature in the thermal conductivity in  experiments by Krishana et.al. \cite{Ong1}

The purpose of this paper is to show that that there is yet another possibility 
to  break time reversal  and parity: {\em spontaneously} in the bulk of the 
d-wave state due to coupling to the impurity spins. We find that: 1) Original 
$d_{x^2-y^2}$ is
{\em unstable} towards formation of the bulk $d_{x^2-y^2}+id_{xy}$ phase  in the
presence of the magnetic impurities under certain conditions.  Time reversal 
symmetry and parity (T and P) are spontaneously broken  as the result of 
interaction between conduction electrons and impurity spins. Below we consider  
magnetic impurities with nonzero spin $S$, such as   Ni with S=1 at low 
temperatures. 
 In the ordered 
phase both the spontaneous magnetization of impurity spins and second component 
of the order parameter are developed simultaneously.   2) We
 construct a Ginzburg-Landau (GL) functional that  contains a linear coupling
between impurity spins and  the  $d_{x^2-y^2}+id_{xy}$ order parameter.
The time reversal violation is natural in this case as it allows the order
parameter to couple directly to the impurity spin. This coupling is possible 
only for $d+id$  and not for $d+is$ symmetry of the order parameter. From the GL description
we find that instability develops as  a {\em second order} phase transition
where both the out-of-plane magnetization and   $d_{xy}$ component
developed together and are proportional to each other. 
3) This instability
occurs at a certain concentration range and is most likely suppressed at
higher concentrations.

Recent  experiments reported the anomaly in the thermal conductivity in
Bi2212 at low temperatures:     the thermal conductivity of
the Bi2212 with Ni  impurities was observed to have a sharp
reduction at $T^*_c = 200 mK$ \cite{Roman1}. These experimental data
indicate the possible superconducting phase transition in the Bi2212  in the 
presence of the
magnetic impurities in a certain concentration range. So far
 the transition has 
been seen only    in the samples with magnetic impurities, e.g.
 Ni as opposed to the nonmagnetic impurities such 
as Zn \cite{Roman1}.  It was reported that the feature in the thermal 
conductivity
 is completely suppressed by applying the 
  field of $H \sim 300 Gauss$. 
The low field and the fact that feature disappears 
is consistent with the superconducting transition
into second phase. 

Our results might   be 
relevant for the experimentally observed transition
  at $T^*_c$ in Bi2212 with 
Ni. 

We will construct a GL functional for the d-wave superconductor with
magnetic impurities.  Ni impurities is just one example of the magnetic
impurities substituting Cu in Cu-O planes. No other assumptions will be made 
about the nature of impurity but the existence of the noncompensated isotropic 
moment   at low temperatures.  We assume that
superconductor is 2D with weak Josephson coupling between layers
in Bi2212. The relevant fields to enter the GL functional are:

$\bullet$ $\Psi_0$ -the order parameter of the original $d_{x^2-y^2}$ component
$\Delta_0 = \Psi_0(\cos k_xa - \cos k_ya)$ ,

$\bullet$ $\Psi_1$- the order
parameter of the $d_{xy}$ component $\Delta_1 = \Psi_1 \sin k_xa \sin k_ya$ 
\cite{com5}, and

$\bullet$ ${\bf M}$- the magnetization of the impurity spins. 

$\Delta_{0,1}$ are the respective gap functions and we
wrote them down explicitly in the momentum space, $a$ is the lattice spacing. 
The impurity spin
magnetization field ${\bf M}(\br) = {e\hbar \over{2mc}}\bS({\bf r})= \sum_{i}
{e\hbar \over{2mc}}\bS(\br_i) f(\br - \br_i )$ with  normalized to unity 
coarse-graining
function $f(\br)$ which has finite range , greater than the superconducting
correlation length $\xi = v_F/\Delta_0$. This step is done    to
introduce a continuum spin field in GL functional \cite{com4}.

We  consider a secondary transition $\Psi_0 \rightarrow  \Psi_0 +
\Psi_1$ at $T^*_c$. The relative phase of  $\Psi_1$ with respect to the
phase of $\Psi_0$  {\em is not fixed} and will be determined by the free
energy minimization.  We will assume that the second transition, if at all,
occurs at $T^*_c \ll T_c$, where $T_c \sim 90 K$ is the first transition
temperature. Hence the order parameter $\Psi_0$, which we can assume to be
real,  is robust and  its free energy $F(\Psi_0)$ can not be expanded in
$\Psi_0$. 

Apart from coupling to the magnetic field ${\bf B}$, which has to be determined 
selfconsistently,  we assume the spin-orbit coupling between the impurity spins 
and orbital moment of the pair:
\beqa
H_{int} = g \hat{L}_z M_z
\label{Hint}
\eeqa
with some coupling constant $g$ \cite{so}. Upon solving the microscopic 
selfconsistency equation one finds that this term generates the coupling between $\Psi_0$ and $\Psi_1$ (see $F_{int}$ below).

Assuming expansion in powers of small  $\bM, \Psi_1$ near second transition, the GL functional  $F=
F_{sc} + F_{magn} + F_{int}$ is:
\beqa
&F_{sc} = F(\Psi_0, {\bf A}) + \alpha_1/2|\Psi_1|^2 + \alpha_2/4 |\Psi_1|^4 + 
\nonumber\\
&\beta |(\nabla_i - 2e/c A_i)\Psi_1|^2, \alpha_{1,2}\geq 0\nonumber\\
&F_{magn} = {a_1\over{2}}|M_z|^2 +  {a_2\over{4}}|M_z|^4 +
{a_3\over{2}}|\nabla_i M_z|^2  + {B^2\over{8\pi}}\nonumber\\
&F_{int} = -{b\over{2i}}(\Psi_0^*\Psi_1 - h.c.)M_z - (M_z + 
{\nu\over{2i}}(\Psi_0^*\Psi_1 - h.c.))B_z 
\label{int}
\eeqa
In writing the coupling between
$M_z$ and $\Psi^*_0\Psi_1$ we need only the term which transforms as
$\Psi_0$ and any function in the
form $\Psi_0 \Phi(|\Psi_0|^2)$   is allowed in the interaction term
\cite{heeb}. We will absorb  $\Phi(|\Psi_0|^2)$ into coupling $b$ in
$F_{int}$, since $\Psi_0$ is constant. The same absorption was done
for the term with coupling $\nu$.  $\Psi_0$ should enter  in $F_{int}$
to satisfy the global $U(1)$ symmetry: free energy should be invariant
under simultaneous $\Psi_{0,1} \rightarrow \Psi_{0,1} \exp(i\theta)$.
$F_{int}$ is similar to linear coupling between magnetic field and the
secondary order parameter, considered by Laughlin \cite{Laugh1}.
Because of strong 2D anisotropy of Bi2212  we consider here 2D
superconductor within the mean field approximation. Weak interplanar
coupling can be accounted for later.

Below we consider  Meissner phase, assuming that there
 is no net ${\bf B}$   and  no mass currents in the bulk.
  This requires ${\bf B} = \nabla_{\bf r} (M_z + {\nu\over{2i}}(\Psi_0^*\Psi_1 - h.c.)) = 0$ and can be satisfied for constant $M_z$ and $L_z$.  There is no 
coupling between $L_z$ and $M_z$ via magnetic field and there is only contact 
spin-orbit coupling. Only first term   in $F_{int}$ is  nonzero therefore.

All but $F_{int}$ terms in the free energy Eq.(\ref{int}) 
are positive and can not produce the instability of the 
original $d_{x^2-y^2}$ state.  $F_{int}$  can be negative 
since it is linear in $\Psi_1$ and $M_z$ and this term is 
crucial in producing second transition. 

In $F_{magn}$ coefficients $a_{1,2}$ are temperature dependent even at low 
temperatures. 
Direct exchange between Ni spins can be
 ignored in low concentration limit when the Ni-Ni distance is 20-30 $\AA$. The 
only
relevant contribution will be
 entropic term in $F_{magn} = E_{magnt} - TS_{magn}$.
 In the presence of
 the ordered moments we get $S_{magn} = S_0 - b n_{imp} M^2_z$ with some 
constant $b$. Hence 
$a_1 = b T n_{imp}$ in $F_{magn}$. Rapid decrease of $a_1$ with temperature 
might explain why second transition
occurs at low temperatures.

Consider $F_{sc}$.   The
second and third terms in  $F_{sc}$  describe be the cost of opening the
fully gapped state with $\Psi_1$ when interaction prefers to keep node, i.e.
pure $d_{x^2-y^2}$ state. The change in free energy due to secondary order
parameter is given by
the difference in  energy of quasiparticles  before and after  $\Psi_1$
component
 is generated. We calculate the change in the energy of the superconductor 
subjected to the {\em homogeneous} external $d_{xy}$ source field: $H_{xy}= 
\kappa \sum_{\bk} \Delta_1(\bk) c^*_{\bk \sigma} c^*_{-\bk -\sigma}$, where $0 < 
\kappa < 1$ is the integration constant and $c_{\bk \sigma}$ -- electron 
operator. Using standard result $\partial_{\kappa} F_{sc} = 1/\kappa \langle 
H_{xy} \rangle$ we find increase of energy at $ \Delta_1 << T << \Delta_0$:
\beqa
&\delta F_{sc}=N_0\int_0^1d\kappa\kappa\int{d\phi\over{2\pi}} |\Delta_1(\phi)|^2 \ln {E_c\over{(|\Delta_0(\phi)|^2 + \kappa^2|\Delta_1(\phi)|^2 + 
T^2)^{1/2}}}\nonumber\\
& ={\alpha_1\over{2}}|\Psi_1|^2+{\alpha_2\over{4}}|\Psi_1|^4\nonumber\\
\label{Fsc}
\eeqa
Here we take, based on the results of microscopic calculation for the induced 
component, that $\Psi_1$ is constant \cite{com1}. $N_0$ is the Density of States at the Fermi energy and $E_c$ is the energy cutoff.

The most interesting is the interaction term $F_{int}$. Few comments are in
order here:

1) The nature
 of the linear  coupling between $M_z$ and $Im\Psi^*_0\Psi_1$ has
 simple physical meaning: for $\Psi_0 + i\Psi_1$ state the orbital part
of the pair wave function has a finite
 component of $L_z = 2$ state:  \beqa \Delta_{\bk \in FS} = \Delta_0 +
i \Delta_1 = \Psi_0 \cos 2\theta + i\Psi_1 \sin2\theta \nonumber\\ =
\Psi_1 exp(2i\theta) + (\Psi_0 - \Psi_1)\cos2\theta \eeqa and
interaction term describes the coupling between the orbital moment
density $L_z \propto Im\Psi^*_0\Psi_1$ of the pair and the spin
magnetization  $M_z$.  One can show the existence of the linear
coupling in microscopic theory by looking at the matrix element of
$H_{int}$    between $\Psi_0$ and $\Psi_1$
states: $F_{int} \propto \langle \Psi_0 \cos 2\theta|-i \hbar
\partial_{\theta} M_z|\Psi_1 \sin 2\theta \rangle$, $\hat{L}_z = -i
\hbar \partial_{\theta}$, $\theta$ is a planar angle.

2) This form of interaction is  possible only  when $\Psi_1$ is the
amplitude of the $d_{xy}$ component. $Im \Psi^*_0\Psi_1$ transforms as
a pseudovector (magnetization) along z-axis. Indeed
 under time reversal T: $\Psi_{0,1} \rightarrow \Psi^*_{0,1}$ and $M_z
\rightarrow -M_z$ so that $F_{int}$ remains invariant. To check spatial
inversion, consider $\pi$ rotation around
 x-axis: $x\rightarrow x, y\rightarrow -y, z\rightarrow -z$. Under this
operation $M_z$
 changes sign, $\Psi_0 \propto d_{x^2-y^2}$ does not change and $\Psi_1
\propto d_{xy}$ does.  Hence the $Im\Psi^*_0\Psi_1$ changes sign and
$F_{int}$ remain invariant. One can check analogously other spatial
transformations to see that indeed $F_{int}$ is a  full scalar. $d+is$ can not 
couple linearly to magnetic field $M_z$.

3) Linear coupling $Im\Psi^*_0\Psi_1 M_z $ enables the second
 phase transition,  where {\em both} $\Psi_1$ and
$M_z$ are developed below some critical temperature $T^*_c$.

4) The only orbital current allowed in the planar geometry generate the
z-axis magnetic moment which can couple only to $M_z$.
 Since  only $F_{int}$ can be negative and drive the transition by coupling to 
$M_z$,  we can ignore other terms and consider free energy as
 a functional of $M_z$ only.

Next, we will minimize the functional Eq.(\ref{int}) and find the
solution for the ordered phase \cite{com3}.
Fix $\Psi_0$ to be real positive and the relative phase of $\Psi_1 =
|\Psi_1|
\exp(i \nu)$ without loss of generality. The set of equations is:
\beqa
&{\partial F\over{\partial M_z}} = a_1 M_z - b \sin \nu \Psi_0\Psi_1 =
0\nonumber\\
&{\partial F\over{\partial \nu}} = - b \cos \nu \Psi_0|\Psi_1| M_z = 
0\nonumber\\
&{\partial F\over{\partial |\Psi_1|}} = \alpha_1 |\Psi_1| +
\alpha_2|\Psi_1|^3 -
 b \Psi_0 M_z \sin \nu = 0\nonumber\\
\label{eqF}
\eeqa

From the second Eq.(\ref{eqF}) it is obvious that $\nu = \pm \pi/2$ and the
phase will be determined by minimization of energy:
\beqa
\nu = \pi/2 \ sgn(b M_z)
\eeqa
This choice takes  the maximum advantage of the $L_zM_z$ coupling and
minimization {\em requires} a complex order parameter $\Psi_0 + i \Psi_1$ in the low
temperature phase. T and P are violated {\em spontaneously}.
Solving the remaining equations we get:
\beqa
& M_z = {b\over{a_1}} \sin \nu \Psi_0|\Psi_1| \nonumber\\
& |\Psi_1|^2 = {1\over{\alpha_2}}({b^2\over{a_1}} \Psi^2_0 - \alpha_1)
 = \chi (T^*_c- T)\nonumber\\
& \delta F = -{\alpha_2\over{4}}|\Psi_1|^4  
\sim  \ |T-T^*_c|^2
\label{sol}
\eeqa
This is the  main result of this paper. Solution Eq.(\ref{sol}) indicates
  that the transition is of the second order with the jump in the 
 specific heat.  We assumed that $|\Psi_0(T)|^2/a_1(T)$ has a  linear
temperature slope near $T^*_c$.  

It follows from the solution Eq(\ref{sol}) that:

1) If coupling $b$ is small enough $b^2\leq \alpha_1 a_1(T=0)/\Psi^2_0(T=0)$
 there is no nontrivial solution for $\Psi_1$.
 For large enough $b$ or small $a_1$ the solution near $T^*_c$ is determined by the
temperature dependent
$\Psi_0(T), a_1(T)$: $\Psi^2_0(T^*_c)/a_1(T^*_c) = \alpha_1/b^2$.
 Strong
temperature and concentration dependence of 
$a_1 \propto n_{imp} T$ might explain why the
second transition occurs at low temperatures: the rapid 
decrease of $a_1$ with  $T$ makes the condition for 
positive solution in Eq(\ref{sol}) possible to satisfy at low $T$.

2) As the function of impurity concentration two effects occur
simultaneously:  condensate density $|\Psi_0|^2$
decreases and $a_1$ increases at increased impurity concentration. This effect 
will lead
eventually to the disappearance of the transition when $\alpha_1$
reaches critical value $\Psi^2_0(T=0) \leq  a_1 \alpha_1/b^2$.
Secondly, the suppression of $\Psi_1$ due to increased impurity
scattering will also lower $T_c^*$. Quick suppression of the transition
temperature $T^*_c$ with impurity concentration should be expected.

3)The above calculation, which we have done in the homogeneous
approximation, is a variational proof of the instability of the
original $d_{x^2-y^2}$ state toward the formation of the $i d_{xy}$
component and spin magnetization simultaneously.  The order 
parameter and magnetization in the  real
system do not have to be homogeneous. It is possible 
that there is inhomogeneous solution with lower energy.
  That is why the homogeneous 
solution
should be considered as a variational approximation to the
 true energy.

4) Strong magnetic field parallel to the layers,
 $ H \gg H_{c1,ab} \sim 1 \ Gauss$ in plane, will 
suppress the second phase.
In the field Ni spins will be aligned in the layers, 
linear coupling term on $H_{int}$ will
be zero and $d_{xy}$ component will vanish. This effect 
might explain the suppression
of the second transition by magnetic field 
$H_{c1,ab} \ll H \leq H_{c1,c} \sim 300 \  Gauss$ , seen in experiment
\cite{Roman1}.

5)  For real systems the interlayer coupling will lead to the true
superconducting instability as opposed to the 2D mean field theory,
considered here. Therefore details of interlayer interactions will be
important in understanding true instability. It could be, for example,
that real transition temperature is strongly determined by interlayer
coupling.

To test the proposed state following experiments can be done. The driving mechanism for the second phase clearly distinguishes between magnetic and nonmagnetic impurities, hence more experiments on Bi2212 with nonmagnetic impurities will be helpful \cite{Roman1}. Because theory predicts the ordering of Ni moments below $T^*_c$, one should be able to detect magnetization in $\mu SR$ experiments. The increased superfluid density due to second component translates into the change in the penetration depth below $T^*_c$ which can be detected. These and other experiments will help to resolve if the proposed mechanism is correct.

An interesting and related case is  when
Ni moments order intrinsically at some $T_m$ : $a_1 \propto T - T_m$.
The linear coupling, discussed here, with necessity will generate
$d_{xy}$ component
at some
temperature. It follows from the solution Eq.(\ref{sol}) with
arbitrarily small $a_1$. First order transition is also possible.

In conclusion, I presented the  mechanism for  a second order phase
transition of original d-wave state into $d+id$ state with
spontaneously broken  T and P.
 In the ordered  phase both impurity spins $M_z$ and $d_{xy}$
component of the order parameter develop and are proportional to each
other. The low temperature phase develops magnetic moment both due to
magnetic impurities spin and because of the finite angular momentum of
$d+id$ state.

I am  grateful to R.B. Laughlin, D.H. Lee, A. Leggett, 
  R. Movshovich, and M. Salkola 
for the useful discussions. This work was supported by US DOE.


\begin{references}
\bibitem{com5} In writing $\Delta_1 = \Psi_1 \sin k_xa \sin k_ya$ we assume that 
$\Psi_1$ is nonzero  everywhere on the Fermi surface.  In principle $d_{xy}$ 
component, will have amplitude $\Psi_1$ and width $w$ near nodes, where $\Psi_1$ 
is finite. Both $\Psi_1$ and $w$ should be determined  from free energy 
minimization.  In our microscopic calculation $\Psi_1$ is generated, for 
example,  by $H_{int} = \hat{L}_z M_z$ coupling and is finite for all angles on 
the Fermi surface, hence we assume $\Psi_1$ to be constant.
\bibitem{com4} For   in-plane impurity concentration  $n_{imp} \sim  1\%$ the 
average distance between impurities becomes of the order of the in-plane 
correlation length $\xi \sim 25 \AA$ and mean field approximation for spin 
magnetization is reasonable.
\bibitem{com1}  In the opposite case of the  $\Psi_1$ component open only in the 
narrow vicinity of the node, within width $w \propto \Psi_1$ the angular 
integral in Eq.(\ref{Fsc}) is restricted to small angles near the nodes.  We 
find in this case that $F_{sc} \propto |\Psi_1|^3 \ln(E_c/max(\Psi_1, T))$ which 
is similar, apart from logarithmic factor,  to the result obtained by Laughlin 
\cite{Laugh1}.
\bibitem{com3} We will focus on the ferromagnetically ordered case. The
realistic configuration of impurity spins can be different and involve
modulations, such as antiferromagnetic order within planes, alternating
order within each plane, etc.   These details depend on particular model one 
choses to
address coupling between spins and conduction electrons. The goal here
is to prove first that there is a variational state with total energy
lower than pure $d_{x^2-y^2}$ state once spins are ordered. The
 lowest energy state might have inhomogeneous spin ordering and
inhomogeneous superconducting order parameter.
\end{references}
\end{document}